\documentclass[aip,apl,10pt,twocolumn,showpacs]{revtex4-1}

\usepackage{amsmath}
\usepackage{graphics}
\usepackage{dcolumn}

\input{MyMc}
\newcommand{\EG}[1]{}
\renewcommand{\new}[1]{#1\xspace}

\begin{document}

\title{Memory cell based on a $\varphi$ Josephson junction}

\author{E. Goldobin}
\author{H. Sickinger}
\affiliation{%
  Physikalisches Institut and Center for Collective Quantum Phenomena in LISA$^+$,
  Universit\"at T\"ubingen, Auf der Morgenstelle 14, D-72076 T\"ubingen, Germany
}

\author{M. Weides}
\affiliation{%
  Peter Gr\"unberg Institute and JARA-Fundamentals of Future Information Technology,
  Forschungszentrum J\"ulich GmbH, 52425 J\"ulich, Germany
}
\altaffiliation{%
  Current address:
  Physikalisches Institut, Karlsruher Institut f\"ur Technologie, 76131 Karlsruhe, Germany
}

\author{N. Ruppelt}
\author{H. Kohlstedt}
\affiliation{%
  Nanoelektronik, Technische Fakult\"at,
  Christian-Albrechts-Universit\"at zu Kiel,
  D-24143 Kiel, Germany
}

\author{R. Kleiner}
\author{D. Koelle}
\affiliation{%
  Physikalisches Institut and Center for Collective Quantum Phenomena in LISA$^+$,
  Universit\"at T\"ubingen, Auf der Morgenstelle 14, D-72076 T\"ubingen, Germany
}

\date{\today} 

\begin{abstract}

The $\varphi$ Josephson junction has a doubly degenerate ground state with the Josephson phases $\pm\varphi$. We demonstrate the use of such a $\varphi$ Josephson junction as a memory cell (classical bit), where writing is done by applying a magnetic field and reading by applying a bias current. In the ``store'' state, the junction does not require any bias or magnetic field, but just needs to stay cooled for permanent storage of the logical bit. Straightforward integration with Rapid Single Flux Quantum logic is possible.

\end{abstract}


\maketitle


Superconducting digital circuits, in particular based on Josephson junctions (JJs), potentially offer much lower energy dissipation per logical operation than the existing semiconducting devices\cite{Mukhanov:2011:ERSFQ}. However any practical computer as well as many  digital devices need compatible superconducting random access memory (RAM). The largest superconducting RAM was demonstrated more than a decade ago and had a capacity of 256 kwords ($1\units{word} \equiv 16\units{bits}$)\cite{Nagasawa:1999:RAM:256kW}. Its operation frequency of $\sim 1\units{GHz}$ is already similar to the one of modern computers and a huge feed current of about $2\units{A}$ was necessary for operation.

Since the beginning of this century the JJ technology progressed significantly. For example, it became possible to obtain JJs with a phase difference of $\pi$ in the ground state\cite{Ryazanov:2001:SFS-PiJJ,Ryazanov:2001:SFS-PiArray,Kontos:2002:SIFS-PiJJ,Oboznov:2006:SFS-Ic(dF),Weides:2006:SIFS-HiJcPiJJ} --- the $\pi$ junctions. $\varphi$ JJs\footnote{We follow the terminology after A. Buzdin\cite{Buzdin:2003:phi-LJJ}} consisting of 0 and $\pi$ parts connected together were proposed\cite{Goldobin:2011:0-pi:H-tunable-CPR} and demonstrated experimentally\cite{Sickinger:2012:varphiExp} recently. Such a JJ has a doubly degenerate ground state with the Josephson phase $\psi=\pm\varphi$, where the value of $0<\varphi<\pi$ depends on design parameters. This results from the peculiar Josephson energy profile $U_J(\psi)$, which has a form of a $2\pi$ periodic double well potential. One can use these two intrinsic states of the $\varphi$ JJ to store one bit of information.

In this paper we demonstrate the operation of a memory cell (1 bit) based on a $\varphi$ JJ --- the so-called ``$\varphi$ bit'', which is the first step towards a \new{$\varphi$} Josephson memory.


For the demonstration of the memory cell we have used the same samples as in Ref.~\onlinecite{Sickinger:2012:varphiExp}. Namely, the sample is a Nb$|$Al$-$AlO$_x$$|$Ni$_{0.6}$Cu$_{0.4}$$|$Nb multilayered heterostructure, with a step-like thickness of the ferromagnetic layer to obtain a 0-$\pi$ JJ of moderate length. The sample fabrication is published elsewhere \cite{Weides:2007:JJ:TaylorBarrier,Weides:2010:SIFS-jc1jc2:Ic(H)}. For this work we have chosen a 0-$\pi$ JJ with an in-plane length $L = 200\units{\mu m}$ and the width $W = 10\units{\mu m}$. By treating the structure as a whole (like a black box with two electrodes), one deals with a $\varphi$ JJ\cite{Goldobin:2011:0-pi:H-tunable-CPR,Lipman:varphiEx,Sickinger:2012:varphiExp}.

\begin{figure*}[!htb]
  \centering\includegraphics{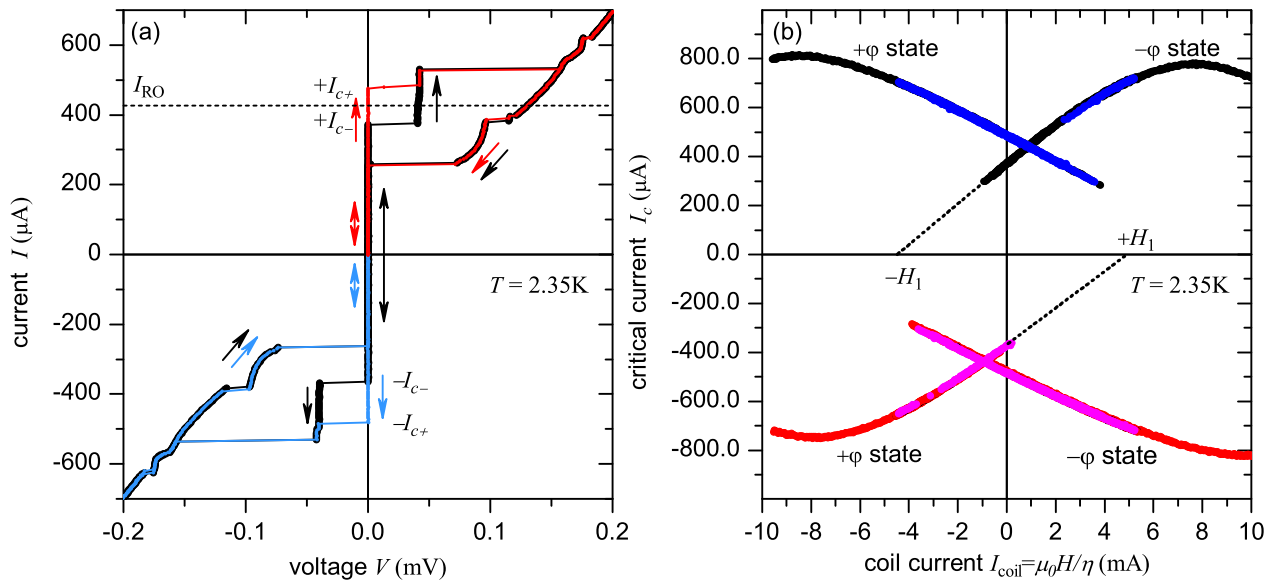}
  \caption{(Color online)
    (a) $V(I)$ characteristic at zero field and (b) $I_c(H)$ curve of the investigated $\varphi$ JJ measured at $T=2.35\units{K}$ (overlap of several measurements with different initial conditions to trace out different branches).
  }
  \label{Fig:IV&IcH}
\end{figure*}

The current-voltage characteristic $V(I)$ as well as the dependence of the critical current $I_c$ on applied in-plane magnetic field $H$ are shown in Fig.~\ref{Fig:IV&IcH}. Already here one can see that we observe two different critical currents, $I_{c-}$ or $I_{c+}$, depending on the initial state of the JJ, $\psi=-\varphi$ or $\psi=+\varphi$, as demonstrated in the original work\cite{Sickinger:2012:varphiExp}. By applying a magnetic field, one can remove this degeneracy and observe only one branch on both $V(I)$ (not shown) and on $I_c(H)$. A point-symmetric $I_c(H)$ with shifted cross-like minimum around zero field, as in Fig.~\ref{Fig:IV&IcH}(b), is the signature of a $\varphi$ JJ.

\begin{figure*}[!tb]
  \centering\includegraphics{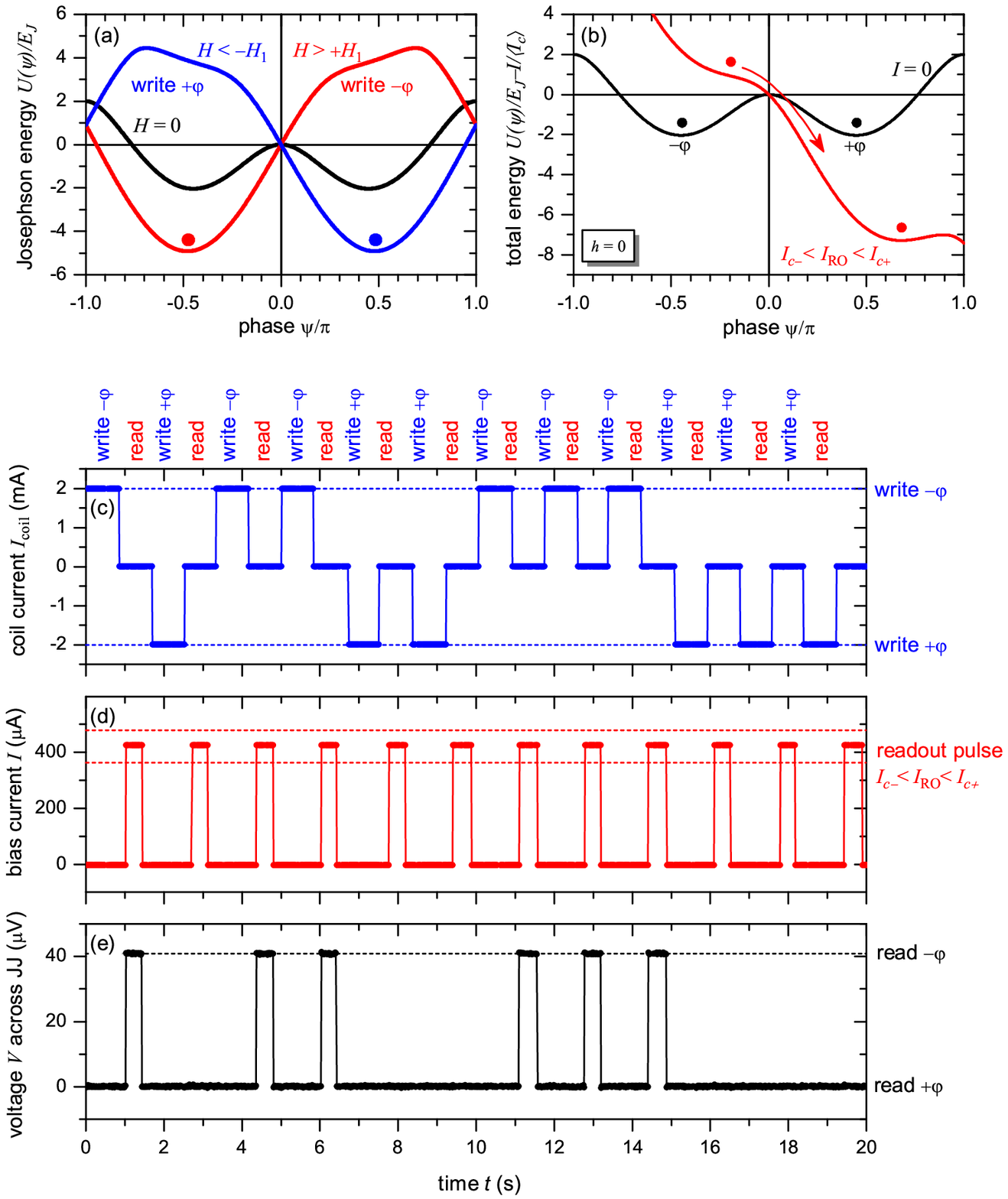}
  \caption{(Color online)
    (a) The principle of writing the state $+\varphi$ or $-\varphi$ by applying a magnetic field.
    (b) The principle of reading the state by applying the bias current $I_{c-}<I_\mathrm{RO}<I_{c+}$.
    (c--e) The sequence of the external magnetic field (c), readout by the bias current (d) and the measured voltage (e), demonstrating the operation of the $\varphi$ JJ as a memory cell.
    The speed of operation is limited by our dc setup (coil) and electronics.
  }
  \label{Fig:WriteRead}
\end{figure*}
%

\new{For writing the state we use the fact that} an applied magnetic field changes the asymmetry between the two wells in the Josephson energy $U(\psi)$ up to the point that one of the wells disappears\cite{Goldobin:2011:0-pi:H-tunable-CPR,Sickinger:2012:varphiExp}, see Fig.~\ref{Fig:WriteRead}(a). Hence one can set (write in) the desired state: $-\varphi$ for $H>H_1$ and $+\varphi$ for $H<-H_1$, where $\pm H_1$ is the field, at which one minimum in $U(\psi)$ disappears. After the state is written, one can return the magnetic field back to zero and the state of the junction will be stored for infinitely long time provided the energy barrier is considerably larger than the energy of thermal or quantum fluctuations.

\new{For reading out the state} we use the fact that the critical current of the $\varphi$ JJ is equal to $I_{c+}$ or $I_{c-}$ depending on the initial state of the JJ, $+\varphi$ or $-\varphi$, see $V(I)$ characteristics shown in Fig.~\ref{Fig:IV&IcH}(a). In our case, $I_{c+}\approx 480\units{\mu A}$ and $I_{c-}\approx 370\units{\mu A}$. Therefore, for the readout we apply a bias current $I_\mathrm{RO}=426\units{\mu A}$ shown in Fig.~\ref{Fig:IV&IcH}(a), which lays in between $I_{c+}$ and $I_{c-}$. Then, if the $\varphi$ JJ switches to a voltage state, its critical current was $I_{c-}<I_\mathrm{RO}$, \ie, the JJ was in the $-\varphi$ state. If, on the other hand, after the application of $I_\mathrm{RO}$ the JJ stays in a zero voltage state, then its critical current is $I_{c+}>I_\mathrm{RO}$, \ie, its is in the $+\varphi$ state.

Note, that this readout technique is destructive, at least when we detect the $-\varphi$ state. After switching to a finite-voltage state, resetting the readout bias current back to zero, the phase may be retrapped either in the $-\varphi$ or in the $+\varphi$ state, depending on the JJ parameters such as on the damping. This retrapping of the phase is investigated in detail elsewhere\cite{Goldobin:Retrap}

In Fig.~\ref{Fig:WriteRead}(c--e) we demonstrate an experimentally recorded sequence of writing and reading $\pm\varphi$ states in different order. We start from an unknown state by applying a positive magnetic field pulse, see Fig.~\ref{Fig:WriteRead}(c), which removes the $+\varphi$ well, see Fig.~\ref{Fig:WriteRead}(a), so that the phase relaxes into the $-\varphi$ well. After such a ``writing'' operation, the magnetic field is reset to zero, Fig.~\ref{Fig:WriteRead}(c). Then we apply a readout pulse, see Fig.~\ref{Fig:WriteRead}(d) and observe a non-zero voltage across the junction, see Fig.~\ref{Fig:WriteRead}(e), indicating a readout of the $-\varphi$ state. During the next write-read cycle, we apply the negative magnetic field $H<0$, see Fig.~\ref{Fig:WriteRead}(c), to change the asymmetry in the opposite direction, see Fig.~\ref{Fig:WriteRead}(a), and write the $+\varphi$ state. After application of the readout pulse, see Fig.~\ref{Fig:WriteRead}(d), we detect no voltage across the junction, see Fig.~\ref{Fig:WriteRead}(e), indicating a $+\varphi$ state indeed. Further, the write-read cycles are repeated for the states $-\varphi, -\varphi, +\varphi,+\varphi,-\varphi,-\varphi,-\varphi,+\varphi,+\varphi,+\varphi$.
One can see that the $\varphi$ bit performs perfectly, reproducing the written sequence. Different longer sequences were played out as well without problems.

We have also investigated the margins of operation of our $\varphi$ bit. The readout current $I_\mathrm{RO}$ should be well in between $I_{c-}$ and $I_{c+}$, \ie, $|I_\mathrm{RO}-I_{c\pm}|\gg I_\mathrm{noise}$ --- the background noise (electronic or thermal) in the bias current circuitry. In our case $I_\mathrm{noise}\sim 3\mu A$, so we have very large margins. The margins in terms of the magnetic field were determined experimentally. We have started seeing faults in operation when the writing magnetic field was decreased down to $|\mu_0 H_\mathrm{write}^\mathrm{min}|\approx 1.9\units{mA}\times\eta$ or increased up to $|\mu_0 H_\mathrm{write}^\mathrm{max}|\approx 2.1\units{mA}\times\eta$, where $\eta\sim 5\units{\mu T/mA}$ is the coil factor. \EG{I find it strange that the interval of $H$ is so small and is about twice smaller than $H_1\sim 4\units{mA}\times\eta$, see Fig.~\ref{Fig:IV&IcH}(b). My explanation is that upon step-like increase of $H$, the coil, having a large inductance, overshoots and the field reaches $H_1$ during the transient process.}


We have demonstrated the use of a $\varphi$ Josephson junction as a memory cell (classical bit). The advantage of such a memory cell is that it is easily controllable by an applied magnetic field. The destructive readout can be performed as described here. However in real circuits, \eg, Rapid Single Flux Quantum (RSFQ) logic, one can construct novel cells that will perform readout with an output in a form of a single RSFQ pulse (or no pulse). In this case, the temperature limitation $T<2.4\units{K}$ necessary in the presented readout scheme to avoid the retrapping of the phase in a $+\varphi$ well upon escape from the $-\varphi$ well, see Fig.~\ref{Fig:WriteRead}(b), is not relevant. By increasing the characteristic voltage $V_c$ one reduces the damping and facilitates operation at $T=4.2\units{K}$. This can be done, \eg, by using  a superconductor with the larger energy gap such as NbN. Besides this, $\lambda_L$ for NbN is larger than for Nb, \ie, $\lambda_J$ will be smaller. In fact, if used in RSFQ circuits, a $\varphi$ JJ is a ready-to-use flip-flop.

In terms of the speed of the operation, the measurements presented in Fig.~\ref{Fig:WriteRead}(c--e) are performed on a 10ms scale (delay time between setting the field/current and reading the voltage). In our case, the field is supplied by a coil, which is very slow and limits the writing time by $\sim5\units{ms}$. Applying the field by an on-chip control line will substantially accelerate the writing cycle. The most effective can be a control line passing through the upper or lower electrode of the JJ. Note also, that the writing time is not limited by the remagnetization (domain flipping or rotation) of the ferromagnetic layer, like in some alternative superconducting memory proposals\cite{Larkin:2012:SIFS-bit}, but rather by a characteristic frequency $\omega_c$ (damping) of the $\varphi$ JJ. For the device presented here $V_c\sim150\units{\mu V}$, \ie, $\omega_c/2\pi\sim75\units{GHz}$. \EG{Note, however, that those are very rough estimates as the RSJ model is not directly applicable here due to the unusual CPR.} In addition, we have a zero-field step at the voltage $V_1 \approx 50\units{\mu V}\propto 1/L$, which appears only sometimes. If one takes shorter JJs, the step will appear at much larger voltages or will disappear, \ie, become irrelevant for $V_1>V_c$. \EG{Ryazanov\cite{Larkin:2012:SIFS-bit} claims $V_c\sim700\units{\mu V}$, but in fact has a gap $V_g=2\units{mV}$ and when JJ is shunted $V_c$ is only $20\ldots30\units{\mu V}$, see their Fig.~1.}

In terms of miniaturization, the $\varphi$ JJ used in this proof-of-principle experiment is rather large ($L=200\units{\mu m}$). However, already for existing technology one can easily reduce its length down to $40$--$50\units{\mu m}$. Further decrease of the size becomes difficult because the margins for the realization of a $\varphi$ junction become very narrow. To realize a $\varphi$ JJ on the basis of a 0-$\pi$ JJ, one has to keep a certain relation between the lengths and critical currents of the 0 and $\pi$ parts\cite{Bulaevskii:0-pi-LJJ,Buzdin:2003:phi-LJJ,Goldobin:2011:0-pi:H-tunable-CPR,Lipman:varphiEx}. If these parts have a normalized length $\sim\lambda_J$ and larger, then these requirements are quite relaxed, \ie, even 30\% of asymmetry still results in a $\varphi$ JJ. On the other hand, if the length of each part is shorter than $\sim\lambda_J$ then these requirements are rather strict (asymmetry should be controlled on a level better than 3\%) and difficult to meet technologically. The way out of this problem is the further development of SIFS 0-$\pi$ junction technology with the aim to increase $j_{c,\pi}$ from the present value of $40\units{A/cm^2}$ up to, say $4\units{kA/cm^2}$ (2 orders of magnitude). \new{This can be done either by using a clean F layer or by using a SISIF structure\cite{Bakurskiy:2013:SIsFS}.} In this case, $\lambda_J\propto1/\sqrt{j_c}$ will become $10$ times smaller, so that one can have a $4\mu m$ large bit. Another possibility to reduce $\lambda_J$ is by increasing the specific inductance of the JJ electrodes
\begin{equation}
  \mu_0 d' \approx
    \lambda_{L}\coth\left( \frac{d_1}{\lambda_{L}} \right)
  + \lambda_{L}\coth\left( \frac{d_2}{\lambda_{L}} \right)
  . \label{Eq:d'}
\end{equation}
This can be done (a) by increasing $\lambda_L$, \eg, by using NbN electrodes; or (b) by decreasing the thicknesses $d_1$ and $d_2$ of the superconducting electrodes.

Alternative $\varphi$ JJ technologies\cite{Pugach:2010:CleanSFS:varphi-JJ,Bakurskiy:2013:S-NF-S:varphi-JJ,Heim:2013:varphi-JJ/S-FN-S}, especially those providing intrinsic $\varphi$ JJs\cite{Gumann:2009:Geometric-varphi-JJ} may, at the end, lead to an extremely high-density memory.

\acknowledgments
We acknowledge financial support by the DFG (via SFB/TRR-21, project A5 as well via project GO-1106/3). H. S. gratefully acknowledges support by the Evangelisches Studienwerk e.V. Villigst.

\bibliography{JJ-digital,JJ,SF,pi,SFS}

\end{document}